\documentclass[aps, prd, amsmath, floats, floatfix, twocolumn, nofootinbib, superscriptaddress, showpacs]{revtex4-1}
\usepackage{graphicx}
\usepackage{color}
\usepackage{latexsym}
\usepackage{amsfonts}

\newcommand{\be}{\begin{eqnarray}}
\newcommand{\ee}{\end{eqnarray}}

\begin{document}

\title{Testing general relativity with the reflection spectrum\\of the supermassive black hole in 1H0707$-$495}

\author{Zheng~Cao}
\affiliation{Center for Field Theory and Particle Physics and Department of Physics, Fudan University, 200433 Shanghai, China}

\author{Sourabh~Nampalliwar}
\affiliation{Center for Field Theory and Particle Physics and Department of Physics, Fudan University, 200433 Shanghai, China}
\affiliation{Theoretical Astrophysics, Eberhard-Karls Universit\"at T\"ubingen, 72076 T\"ubingen, Germany}

\author{Cosimo~Bambi}
\email[Corresponding author: ]{bambi@fudan.edu.cn}
\affiliation{Center for Field Theory and Particle Physics and Department of Physics, Fudan University, 200433 Shanghai, China}
\affiliation{Theoretical Astrophysics, Eberhard-Karls Universit\"at T\"ubingen, 72076 T\"ubingen, Germany}

\author{Thomas~Dauser}
\affiliation{Remeis Observatory \& ECAP, Universit\"at Erlangen-N\"urnberg, 96049 Bamberg, Germany}

\author{Javier~A.~Garc\'ia}
\affiliation{Cahill Center for Astronomy and Astrophysics, California Institute of Technology, Pasadena, CA 91125, United States}
\affiliation{Remeis Observatory \& ECAP, Universit\"at Erlangen-N\"urnberg, 96049 Bamberg, Germany}
\affiliation{Harvard-Smithsonian Center for Astrophysics, Cambridge, MA 02138, United States}

\date{\today}

\begin{abstract}
Recently, we have extended the X-ray reflection model {\sc relxill} to test the spacetime metric in the strong gravitational field of astrophysical black holes. In the present Letter, we employ this extended model to analyze \textsl{XMM-Newton}, \textsl{NuSTAR}, and \textsl{Swift} data of the supermassive black hole in 1H0707$-$495 and test deviations from a Kerr metric parametrized by the Johannsen deformation parameter $\alpha_{13}$. Our results are consistent with the hypothesis that the spacetime metric around the black hole in 1H0707$-$495 is described by the Kerr solution.
\end{abstract}

\maketitle

%%%%%%%%%%%%%%%%%%%%%%%%%%%%%%%

In 4-dimensional general relativity, the no-hair theorem guarantees that the only stationary and asymptotically flat vacuum black hole solution, which is regular on and outside the event horizon, is the Kerr metric~\cite{hair}. It is also remarkable that the spacetime around astrophysical black holes formed by complete gravitational collapse in the Universe should be well approximated by the Kerr geometry~\cite{book}. Nevertheless, general relativity has been mainly tested in weak gravitational fields, in particular with Solar System experiments and radio observations of binary pulsars~\cite{will}. The strong gravity regime is largely unexplored, and there are a number of scenarios beyond Einstein's gravity that provide the same predictions for weak fields and present differences when gravity becomes strong.

The study of the properties of the electromagnetic radiation emitted by the gas in the accretion disk can potentially probe the spacetime metric around astrophysical black holes and test the Kerr nature of these objects~\cite{review}. Previous work has shown that X-ray reflection spectroscopy (the so-called iron line method)~\cite{ironkerr} is a promising technique to do this job~\cite{iron}. Currently, the most advanced X-ray reflection model to describe the spectrum from the strong gravity region of a Kerr black hole is {\sc relxill}~\cite{relxill}. In Ref.~\cite{relxill_nk}, we have described {\sc relxill\_nk}, an extension of {\sc relxill} to non-Kerr spacetimes (here {\sc nk} stands for Non-Kerr), and we have shown with some simulations how this new model can test the nature of astrophysical black holes. In this Letter, we employ {\sc relxill\_nk} for the first time to analyze real data and constrain possible deviations from the Kerr solution.

Let us first briefly review the physics and astrophysics behind X-ray reflection spectroscopy. Within the disk-corona model~\cite{corona}, an accreting black hole is surrounded by a geometrically thin and optically thick disk. The corona is a hotter cloud near the black hole. For instance, it might be the base of the jet, the atmosphere above the inner part of the disk, or some accreting material between the disk and the black hole. Its geometry is currently unknown. Because of inverse Compton scattering of thermal photons from the disk off free electrons in the corona, the latter becomes an X-ray source with a power-law spectrum. The corona also illuminates the disk, producing a reflection component with some fluorescent emission features, the most prominent of which is usually the iron K$\alpha$ line, which is at 6.4~keV in the case of neutral and weakly ionized iron and shifts up to 6.97~keV for H-like iron ions. Due to gravitational redshift, Doppler boosting, and light bending, the reflection spectrum is detected in the flat faraway region with a shape different from that at the emission point, and encodes all the details about the strong gravity region near the black hole~\cite{review}.

There are two natural approaches to test the Kerr black hole hypothesis~\cite{book}. In the so-called top-down approach, we consider a specific alternative theory of gravity in which black holes are not described by the Kerr metric and we check whether astrophysical data prefer the Kerr or non-Kerr metric. There are two problems to following this method. First, there are a large number of alternative theories of gravity, and none seems to be more motivated than others, so we should repeat the analysis for every theory. Second, rotating black hole solutions in alternative theories of gravity are known only in quite exceptional cases, while the non-rotating or slow-rotating solutions are not very useful to test astrophysical black holes because the spin plays an important role in the shape of the spectrum.

In the bottom-up approach, we employ a phenomenological test-metric in which possible deviations from the Kerr solution are quantified by one or more ``deformation parameters''. The Kerr metric is recovered when all the deformation parameters vanish, and we want to check whether astrophysical data require vanishing deformation parameters; that is, if astrophysical black holes are indeed the Kerr black holes as expected in Einstein's gravity. There are several such phenomenological metrics available in literature today~\cite{review}. It is important to note that these metrics are not always obtained from some alternative theories of gravity. As such, these metrics and their deformation parameters do not have a well motivated background. Their significance instead lies in the fact that they capture deviations from a Kerr metric.

In this Letter, as an explorative study, we follow this bottom-up approach and we employ the Johannsen metric with the deformation parameter $\alpha_{13}$~\cite{j-m} (see also the Supplemental Material for the line element and the physical interpretation). The Kerr metric is recovered when $\alpha_{13} = 0$. In order to have a regular exterior region (no singularities or closed time-like curves), we have to impose the following restriction to the value of $\alpha_{13}$~\cite{j-m}
\be\label{eq-bound}
\alpha_{13} \ge - \left( 1 + \sqrt{1 - a^2_*} \right)^3 \, ,
\ee
where $a_*$ is the dimensionless spin parameter. We choose to perform the analysis with $\alpha_{13}$ here as an illustration of the capabilities of this model for testing non-Kerr metrics. Further studies, involving both $\alpha_{13}$ and $\alpha_{22}$ and other astrophysical sources, are underway.

The supermassive black hole in the Narrow Line Seyfert~1 galaxy 1H0707$-$495 looks to be a quite promising source for testing the Kerr metric using X-ray reflection spectroscopy. Its spectrum has significant edge features, which are commonly interpreted as an extremely strong reflection component. Previous studies that assumed the Kerr metric and a reflection dominated spectrum found the inner edge of the accretion disk very close to the black hole (thus increasing the relativistic effects in the spectrum),  a moderate inclination angle, and an extremely high iron abundance~\cite{0707,fabian11,0707td12,kara15}. Note that some authors suggest that the spectrum is instead dominated by a powerful wind. This is clear in IRAS13224~\cite{parker}, and it may be possible in 1H0707$-$495 as well~\cite{hagino}. In this Letter we have focused on the most popular reflection model~\cite{0707,fabian11,0707td12,kara15}, because our main motivation is to test the new model rather than to determine which is the correct model for this source, but the wind model is also an important scenario and it should be investigated in a more detailed study.

\textsl{XMM-Newton}, \textsl{NuSTAR}, and \textsl{Swift} observations of 1H0707$-$495 are shown in Tab.~\ref{t-obs}. In our study, for \textsl{XMM-Newton} we have only considered the observation in 2011: it corresponds to the lowest flux state ever observed possessing clear edge features. For the same reason, the 2011 observation has been investigated by several authors, which is helpful for the choice of the models and the comparison of the results. The three separated observations of \textsl{NuSTAR} in 2014 have simultaneous snapshots of \textsl{Swift}. However, the second \textsl{Swift} observation was taken during an anomaly period of this mission and therefore was not included in our analysis. A brief description of the data reduction is reported in the Supplemental Material.

\begin{table}[h]
\centering
\begin{tabular}{ccccccc}
\hline
Mission & \hspace{0.2cm} & Obs.~ID & \hspace{0.2cm} & Year & \hspace{0.2cm} & Exposure (ks) \\
\hline
\textsl{XMM-Newton} && 0511580101 && 2008 && 124 \\
 && 0511580201 && 2008 && 124 \\
 && 0511580301 && 2008 && 123 \\
 && 0511580401 && 2008 && 122 \\
 && 0653510301 && 2010 && 117 \\
 && 0653510401 && 2010 && 128 \\
 && 0653510501 && 2010 && 128 \\
 && 0653510601 && 2010 && 129 \\
 && 0554710801 && 2011 && 98 \\
\hline
 \textsl{NuSTAR} && 60001102002 && 2014 && 144 \\
 && 60001102004 && 2014 && 49 \\
 && 60001102006 && 2014 && 47 \\
\hline
 \textsl{Swift} && 00080720001 && 2014 && 20 \\
 && 00080720003 && 2014 && 17 \\
 && 00080720004 && 2014 && 17 \\
\hline
\end{tabular}
\caption{Observations of 1H0707$-$495. In this Letter, we have only considered the \textsl{XMM-Newton} observation of 2011, the three \textsl{NuSTAR} observations, and the first and the third \textsl{Swift} observations. \label{t-obs}}
\end{table}

\begin{table*}[t]
\centering
\begin{tabular}{lcccccc}
\hline\hline
\hspace{1.5cm} & Analysis~1 & \hspace{0.3cm} & Analysis~2 & \hspace{0.3cm} & Analysis~3 & \hspace{2.0cm} \\
\hline
Data & \textsl{XMM-Newton}~2011 && \textsl{XMM-Newton}~2011 && \textsl{NuSTAR}+\textsl{Swift} & \\
Model & 1 && 2 && 3 & \\
\hline
$a_*$ & $0.96^{+0.01}_{-0.08}$ && $>0.98$ && $>0.99$ & {\sc relxill\_nk} \\
$\alpha_{13}$ & $-0.8^{+1.4}_{-0.7}$ && $-0.05^{+0.1}_{-1.0}$ && $-0.6^{+0.6}_{-0.2}$ & \\
$i$ [deg] & $38^{+4}_{-7}$ && $49^{+2}_{-2}$ && $41^{+2}_{-3}$ & \\
$q$ & $3.6^{+1.1}_{-0.4}$ && $3.9^{+0.5}_{-0.5}$ && $3.7^{+0.1}_{-0.1}$ & \\
$\Gamma$ & $1.33^{+0.07}_{-0.10}$ && $2.49^{+0.03}_{-0.02}$ && $3.29^{+0.02}_{-0.01}$ & \\
&&&&& $2.59^{+0.04}_{-0.02}$ & \\
&&&&& $3.13^{+0.06}_{-0.01}$ & \\
$\log\xi$ & $< 1.79$ && $1.29^{+0.02}_{-0.05}$ && $2.15^{+0.24}_{-0.07}$ & \\
$A_{\rm Fe}$ & $> 8.6$ && $>9.3$ && $>9.6$ & \\
\hline
$T_{\rm in}$ & $0.150 \pm 0.003$ &&&&& {\sc diskbb} \\
\hline
$\log\xi'$ &&& $3.30^{+0.01}_{-0.06}$ &&& {\sc relxill\_nk} \\
$A_{\rm Fe}'$ &&& $>9.6$ &&& \\
\hline
$\chi^2$/dof & $127/94=1.35$ && $157/94=1.67$ && 1938/3246 (C-stat/dof) & \\
\hline\hline
\end{tabular}
\caption{Summary of the best-fit values. The row Data indicates which observations have been used. The row Model indicates the Xspec model employed, and the number refers to that in Eq.~(\ref{eq-m}). The reported uncertainty corresponds to the 90\% confidence level. In Analysis~3, we have used the Cash-statistics instead of the $\chi^2$ one. See the text for more details. \label{t-fit}}
\end{table*}

\begin{figure*}[t]
\begin{center}
\includegraphics[type=pdf,ext=.pdf,read=.pdf,width=8.7cm]{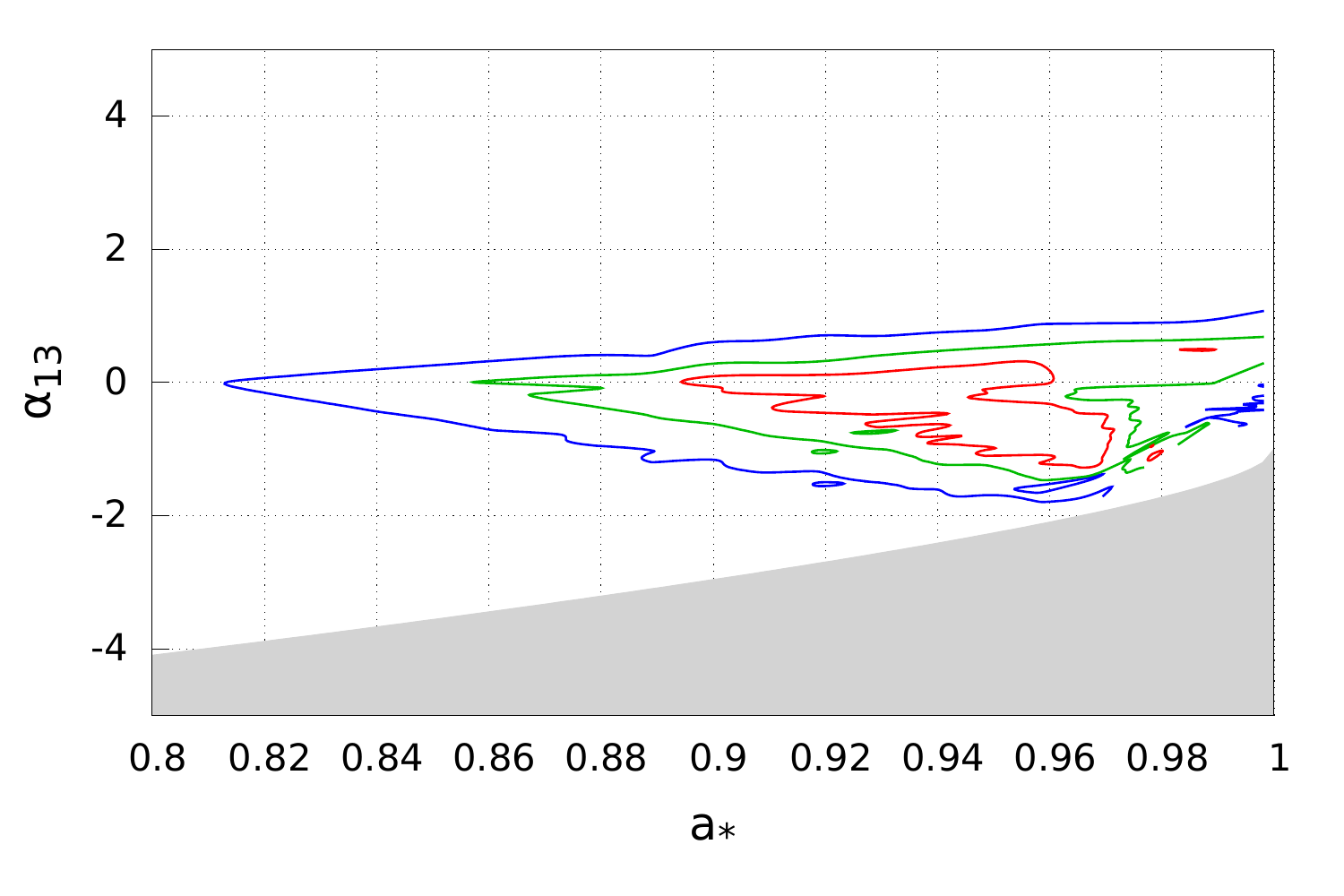}
\includegraphics[type=pdf,ext=.pdf,read=.pdf,width=8.7cm]{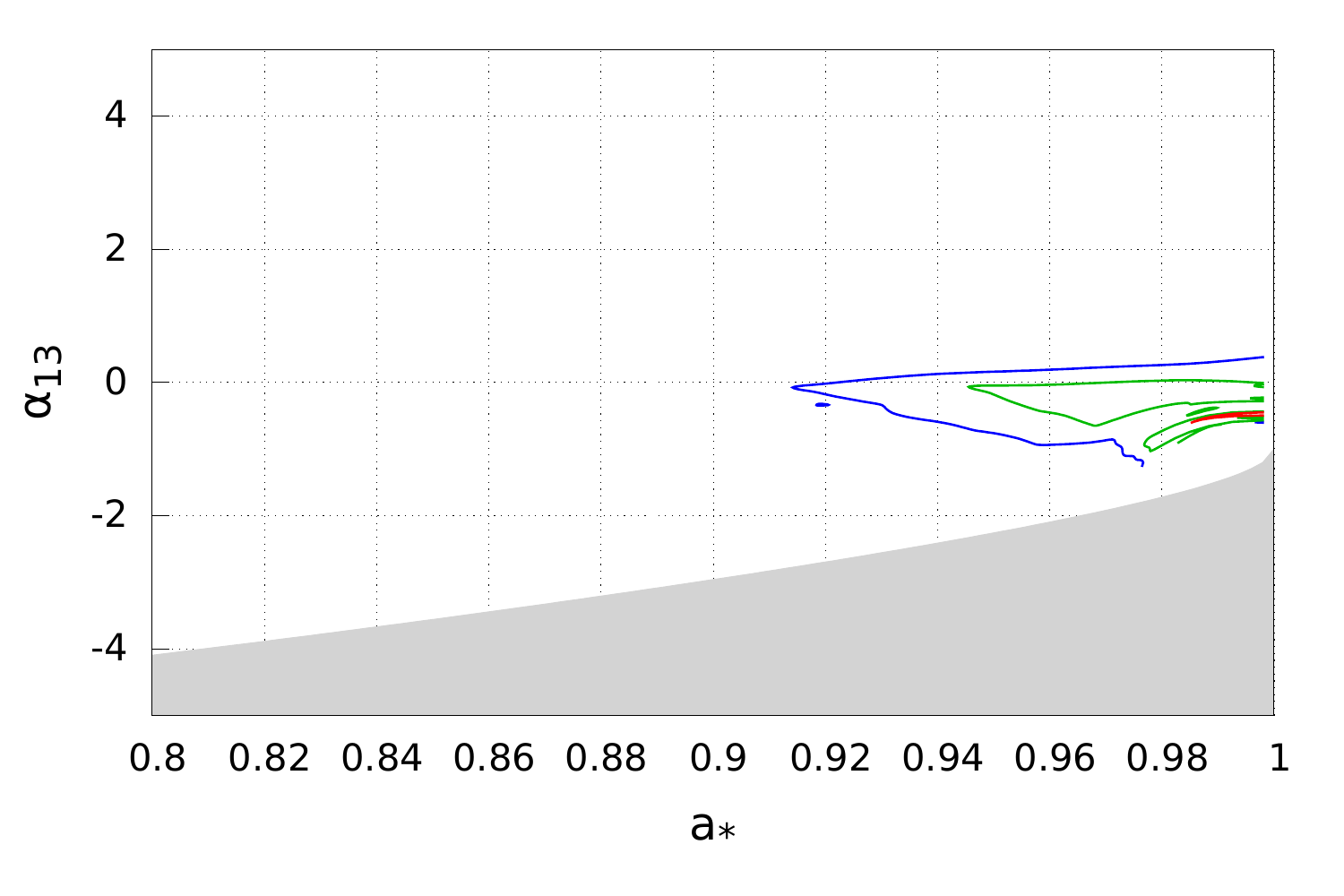}
\end{center}
\vspace{-0.5cm}
\caption{Constraints on the spin parameter $a_*$ and the Johannsen deformation parameter $\alpha_{13}$ from the \textsl{XMM-Newton} data of 2011: Analysis~1 (left panel) and Analysis~2 (right panel). The red, green, and blue lines indicate, respectively, the 68\%, 90\%, and 99\% confidence level curves for two relevant parameters. The grayed region is ignored in our study because it does not meet the condition in Eq.~(\ref{eq-bound}). See the text for more details. \label{f-xmm}}
\end{figure*}

\begin{figure}[h]
\begin{center}
\includegraphics[type=pdf,ext=.pdf,read=.pdf,width=8.7cm]{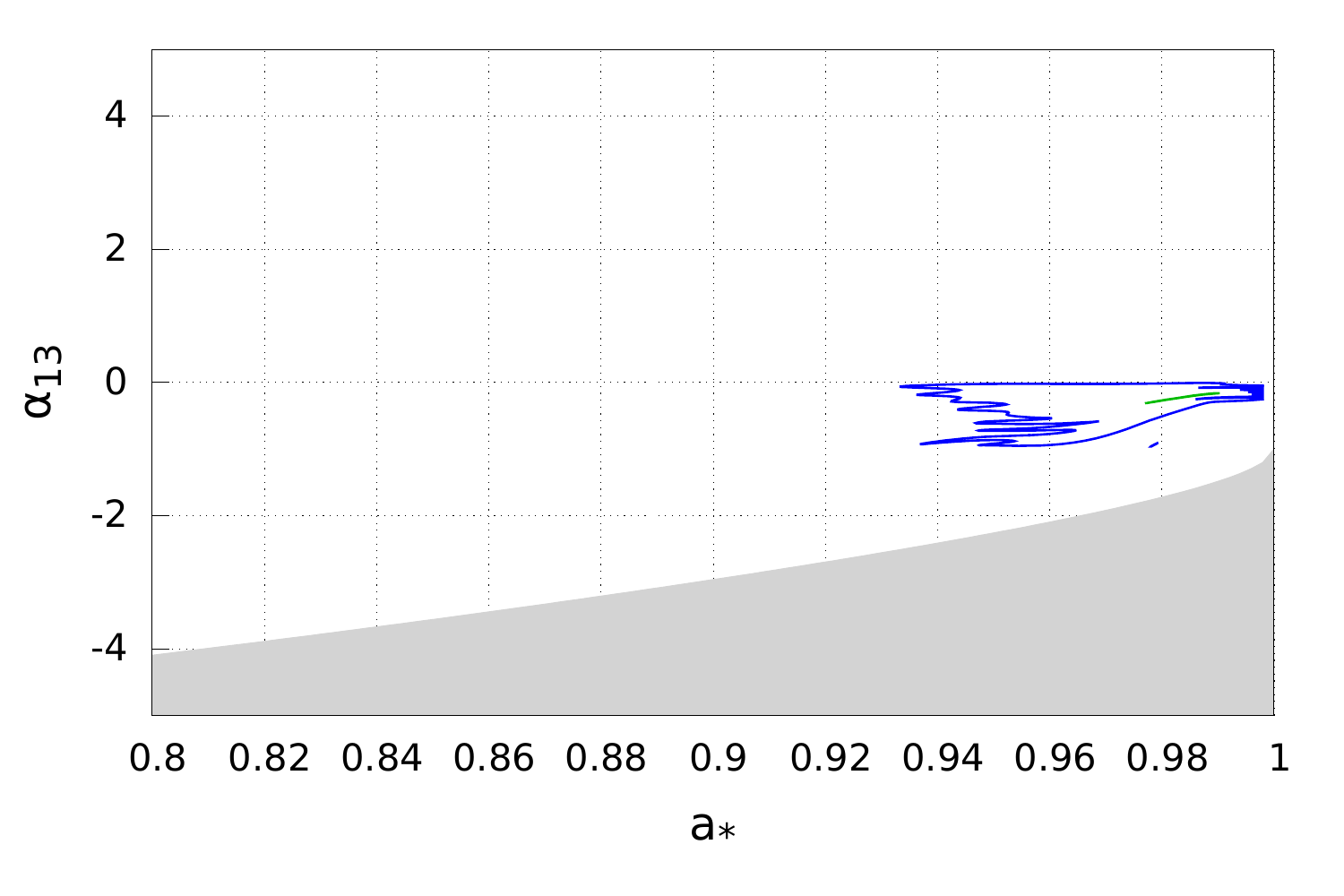}
\end{center}
\vspace{-0.5cm}
\caption{As in Fig.~\ref{f-xmm} from Analysis~3 (\textsl{NuSTAR}+\textsl{Swift}). The 68\% confidence level curve is too thin to be plotted. See the text for more details. \label{f-nustar}}
\end{figure}

We have performed three separated studies (named Analysis~1, 2, and 3) employing the following models respectively
\be\label{eq-m}
&&\text{Model~1: {\sc TBabs*(relxill\_nk+diskbb)}} \, , \nonumber\\
&&\text{Model~2: {\sc TBabs*(relxill\_nk+relxill\_nk)}} \, , \nonumber\\
&&\text{Model~3: {\sc TBabs*relxill\_nk}} \, . 
\ee
Our results are summarized in Tab.~\ref{t-fit} and in Figs.~\ref{f-xmm} and \ref{f-nustar}. The reflection spectrum of the disk in the Johannsen metric is described by {\sc relxill\_nk}, in which the free parameters are the black hole spin $a_*$, the deformation parameter $\alpha_{13}$, the  inclination angle of the disk $i$, the emissivity index $q$ assuming a simple power-law $1/r^q$ where $r$ is the radial coordinate, the photon index of the primary component from the corona $\Gamma$, the ionization of the disk $\log\xi$, and the iron abundance $A_{\rm Fe}$ (in Solar units). {\sc TBabs} takes the galactic dust absorption into account and the column number density has been set to the value measured for 1H0707$-$495 ($N_{\rm H} = 5.8 \cdot 10^{20}$~cm$^{-2}$)~\cite{kara15}.

In Analysis~1, we have fitted the \textsl{XMM-Newton} EPIC-pn data with Model~1. We have employed {\sc diskbb} to fit the ``soft excess'' around 1~keV as done in Ref.~\cite{fabian11}. The corresponding constraint on the spin and the deformation parameters are shown in the left panel in Fig.~\ref{f-xmm}, where the red, green, and blue lines indicate, respectively, the 68\%, 90\%, and 99\% confidence level curves.

In Analysis~2, we have fitted the same \textsl{XMM-Newton} data with Model~2 following what was done with a Kerr metric in Refs.~\cite{fabian11,kara15}. The double reflection model is one of the most popular models to fit the soft excess in AGN like 1H0707$-$495 in which the soft spectrum changes significantly with the flux state. There are a few physical scenarios to motivate a double reflection model, but the basic idea is that there are certain inhomogeneities in the accretion disk. For instance, the density of the disk photosphere may be patchy, leading to mixed regions of high and low ionization~\cite{lohfink}; the surface of the disk may have regions of different density~\cite{fabian11}; it is possible that we are looking at a disk with different layers~\cite{kara15}. The parameters of the two reflection models are tied with the exception of the ionization, the iron abundance, and the normalization. The corresponding constraints on $a_*$ and $\alpha_{13}$ are shown in the right panel in Fig.~\ref{f-xmm}.

The minimum of the reduced $\chi^2$ is not very close to 1 for Analysis~1 and Analysis~2. This is because the \textsl{XMM-Newton} data have a very high signal to noise ratio below 1.5~keV and any model that cannot perfectly fit the soft excess has a relatively large reduced $\chi^2$ (see Ref.~\cite{wilkins14} for a discussion on this point). The residuals between 1 and 4~keV might be due, for example, to a highly ionized outflowing wind~\cite{0707td12}. We could obtain a reduced $\chi^2$ closer to 1 excluding data below 1 or 1.5~keV, as done in some of previous studies in the literature~\cite{wilkins14}. Note, however, that the difficulty to fit the soft energy band is not crucial in the present study, for which the goal is to test the Kerr metric and constrain on the deformation parameter $\alpha_{13}$, because our results are mainly sensitive to the iron K$\alpha$ complex, which is at higher energies. Note that the fit is driven by the small error bars from the soft energy band.

Finally, in Analysis~3 we have fitted the \textsl{NuSTAR} and \textsl{Swift} data with Model~3 following the study in the Kerr metric in Ref.~\cite{kara15}. We impose that the values of the model parameters are the same for the three observations, with the exception of the photon index $\Gamma$, as done in~\cite{kara15} (in Analyses~1 and 2 we have only one photon index because we only consider the observation of 2011). Note that in Analysis~3 we have used the Cash-statistics because of low photon count. The constraints on $a_*$ and $\alpha_{13}$ are shown in Fig.~\ref{f-nustar}; they are better than those from XMM-Newton, but it is not easy to identify the main reason, as the observations are different (source at different times, different exposure times, different energy range of the data, etc.).

In Ref.~\cite{relxill_nk}, generic simulations were performed to test the capabilities of {\sc relxill\_nk} in analyzing observations from {\it present} and {\it future} instruments. We found that \textsl{LAD/eXTP}~\cite{snzhang} can provide significantly stronger constraints on $\alpha_{13}$ than \textsl{NuSTAR}. Here we consider the specific case of 1H0707$-$495 and hypothetical future observations with \textsl{X-IFU/Athena}~\cite{athena}. For Model~1, which is less constrained, a 300~ks observation can distinguish a spacetime with $\alpha_{13} = - 0.5$ from Kerr solutions at 99\% confidence level (see Supplemental Material for more details). Thus, 1H0707$-$495 with a best-fit $\alpha_{13}$ of -0.8, if observed with \textsl{X-IFU/Athena} for 100~ks, will be clearly distinguishable from a Kerr black hole. On the other hand, if $\alpha_{13} = 0.5$, it is not possible to exclude Kerr and $\alpha_{13} = -1$ even with an observation of 300~ks. The constraint on $\alpha_{13}$ strongly depends on the black hole spin. An extended study of the constraining power of present and future X-ray missions will be presented in a future paper.

{\it Conclusions ---} 
In this Letter, we have employed for the first time a new version of {\sc relxill} designed to test the Kerr nature of astrophysical black holes to analyze \textsl{XMM-Newton}, \textsl{NuSTAR}, and \textsl{Swift} data of the supermassive black hole in 1H0707$-$495. We have chosen this source because the spectrum has a very strong iron K$\alpha$ line and the inner edge of the accretion disk extends up to very small radii. Assuming that the spectrum is reflection dominated, our results are summarized in Tab.~\ref{t-fit} and in Figs.~\ref{f-xmm} and \ref{f-nustar}, and are consistent with the assumption that the metric around the supermassive black hole in 1H0707$-$495 is described by the Kerr solution, as expected in general relativity. Work is currently underway to study other black holes with {\sc relxill\_nk} as well as to constrain other deformation parameters or to test black hole metrics from specific gravity theories.

%%%%%%%%%%%%%%%%%%%%%%%%%%%%%%%

{\it Acknowledgments ---}
Z.C. thanks the Cahill Center for Astronomy and Astrophysics at Caltech for hospitality during his visit where part of this work was done. The work of Z.C., S.N., and C.B. was supported by the National Natural Science Foundation of China (Grant No.~U1531117) and Fudan University (Grant No.~IDH1512060). C.B. and J.A.G. acknowledge support from the Alexander von Humboldt Foundation.

%%%%%%%%%%%%%%%%%%%%%%%%%%%%%%%

\newpage

\section*{Supplemental Material}

\subsection*{Johannsen metric with the deformation parameter~$\alpha_{13}$}

In Boyer-Lindquist coordinates, the line element of the Johannsen metric with the deformation parameter $\alpha_{13}$ reads (we use units in which $G_{\rm N} = c = 1$)~\cite{j-m}
\be
ds^2 &=& - \frac{\Sigma \left(\Sigma - 2 M r \right)}{A^2} \, dt^2
+ \frac{\Sigma}{\Delta} \, dr^2 + \Sigma \, d\theta^2
\nonumber\\ &&
+ \frac{\left[ \left(r^2 + a^2\right)^2 \left(1 + \delta\right)^2 
- a^2 \Delta \sin^2\theta\right] 
\Sigma \sin^2\theta}{A^2} \, d\phi^2
\nonumber\\ 
&& - \frac{2 a \left[ 2 M r + \delta \left(r^2 + a^2\right) \right] 
\Sigma \sin^2\theta}{A^2} \, dt \, d\phi \, ,
\ee
where $a = J/M$, $\Sigma = r^2 + a^2 \cos^2\theta$, $\Delta = r^2 - 2 M r + a^2$, and
\be
A = \Sigma + \delta \left(r^2 + a^2\right) \, , \quad
\delta = \alpha_{13} \left(\frac{M}{r}\right)^3 \, .
\ee
The Kerr metric is recovered when $\alpha_{13} = 0$. In order to have a regular exterior region (no singularities or closed time-like curves), we have to impose the following restriction to the value of $\alpha_{13}$~\cite{j-m}
\be\label{eq-bound}
\alpha_{13} \ge - \left( 1 + \sqrt{1 - a^2_*} \right)^3 \, ,
\ee
where $a_* = a/M$ is the dimensionless spin parameter.

$\alpha_{13}$ alters the metric coefficients $g_{tt}$, $g_{t\phi}$, and $g_{\phi\phi}$. It mixes with $a_*$ in a non-trivial way, but generally speaking $\alpha_{13} < 0$ ($> 0$) makes the gravitational field weaker (stronger) and therefore moves the radius of the innermost stable circular orbit (ISCO) to smaller (larger) values. From simulations published in our previous work~\cite{relxill_nk}, we expect that $\alpha_{13}$ and $\alpha_{22}$ have a relatively strong impact on the reflection spectrum, the impact of $\epsilon_3$ is weaker, and that of $\alpha_{52}$ is much weaker.

\subsection*{\textsl{XMM-Newton} data}

In the analysis of the 2011 \textsl{XMM-Newton} observation, we have only used the EPIC-pn data for simplicity. All data files were processed using XMM-Newton Science Analysis System SAS v16.0.0 and the current calibration files CCF updated to 14~March~2017. The spectra and light-curves were extracted using the tool {\it evselect} with default pattern. The source spectra were extracted from a circular region of radius of 35~arcsec and the background region was on the same chip. The effective area and redistribution matrix were produced by {\it arfgen} and {\it rmfgen}, respectively. EPIC-pn was in full window mode during the observation and no evidence of pile-up is found. All EPIC-pn spectra were then binned to a minimum of 20 photon counts per bin before analysis.

\begin{figure*}[h]
\begin{center}
\includegraphics[type=pdf,ext=.pdf,read=.pdf,width=9.3cm]{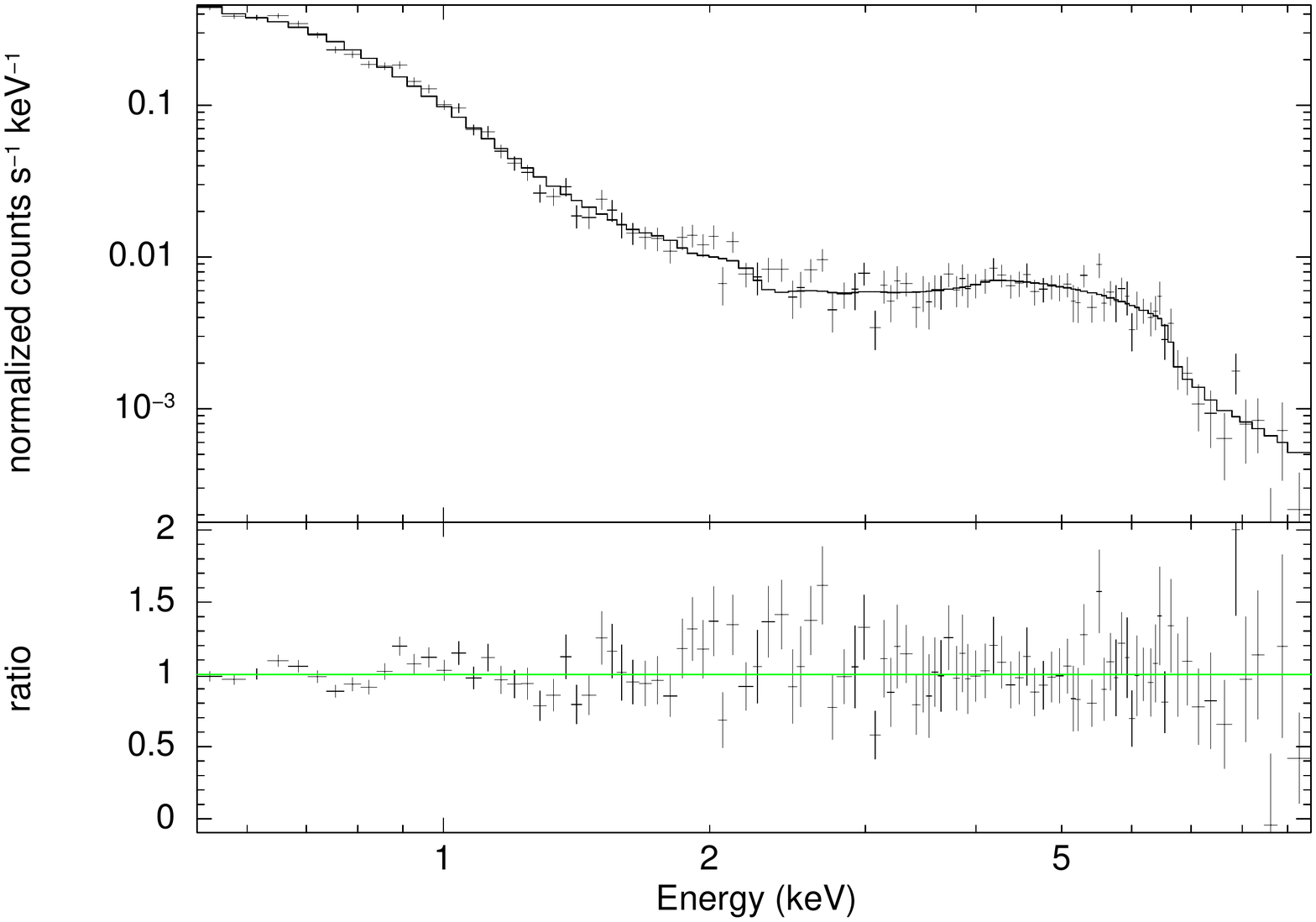}
\hspace{-1.0cm}
\includegraphics[type=pdf,ext=.pdf,read=.pdf,width=9.3cm]{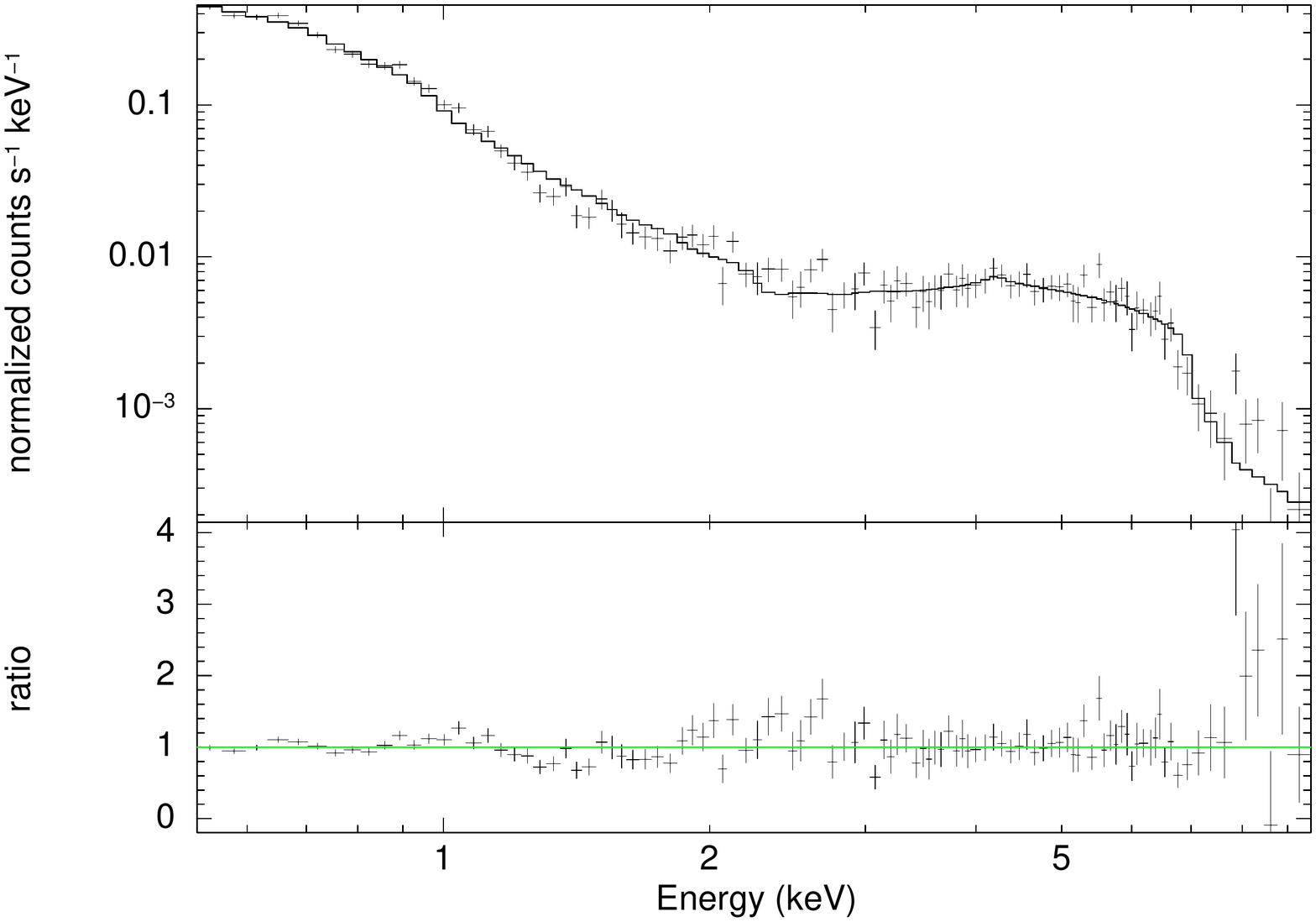}
\end{center}
\vspace{-0.2cm}
\caption{Data and data-to-model ratio of the \textsl{XMM-Newton} observation of 2011: Analysis~1 (left panel) and Analysis~2 (right panel). \label{fs-xmm}}
\begin{center}
\includegraphics[type=pdf,ext=.pdf,read=.pdf,width=9.5cm]{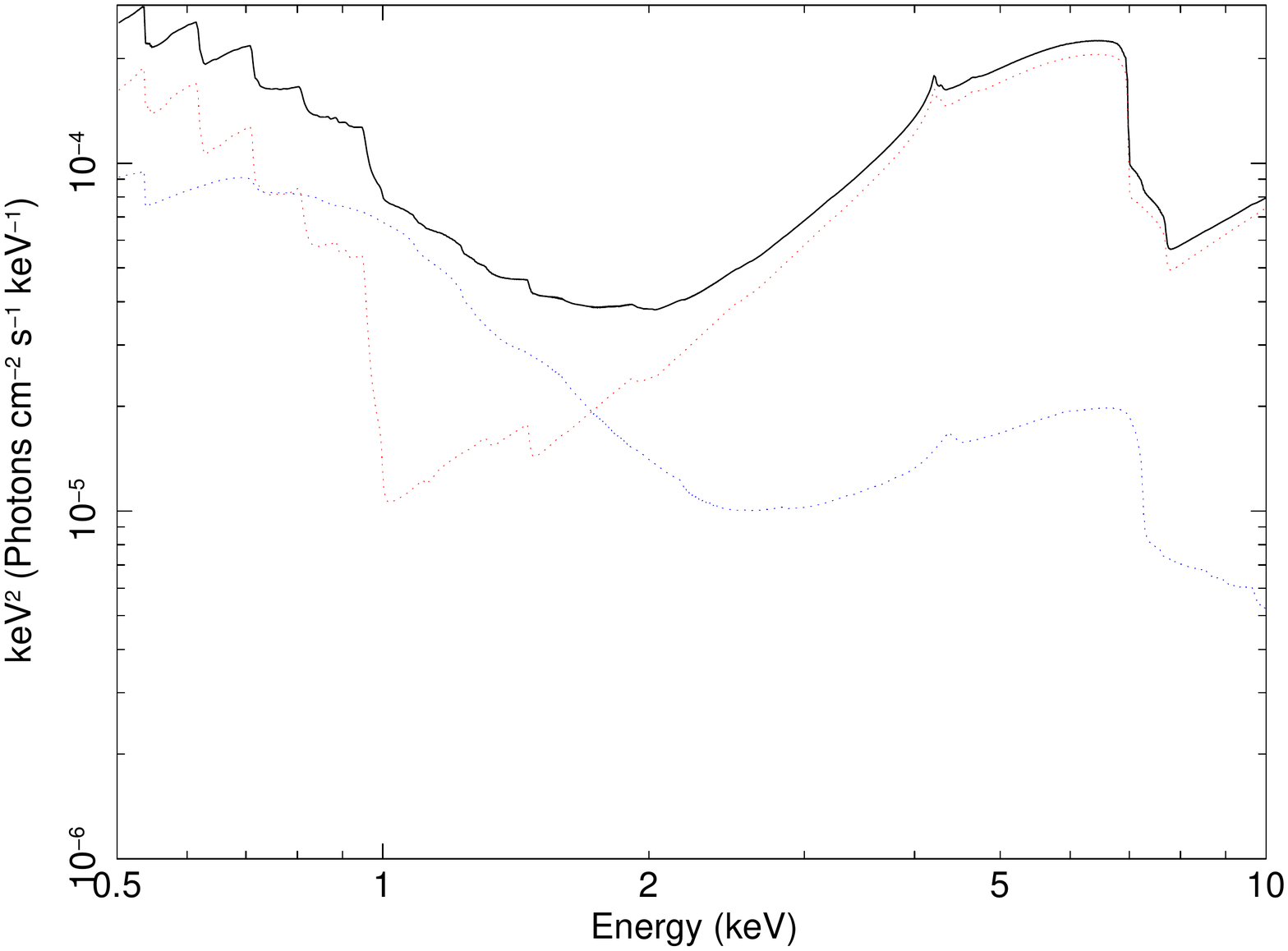}
\end{center}
\vspace{-0.2cm}
\caption{${\rm energy^2 \cdot model}$ for Analysis~2. The red and blue lines are the reflection components with, respectively, lower and higher value of $\xi$. The black line is the whole model. \label{fs-rreemod}}
\end{figure*}

Fig.~\ref{fs-xmm} shows the data and the data to model ratio of Analysis~1 (left panel) and Analysis~2 (right panel). While the minimum of the reduced $\chi^2$ is not very close to 1, the plots of the data to model ratio show that the fits are good. Note that the fits are driven by the small error bars form the soft energy band, but we know that the strong iron K$\alpha$ edge is there and therefore it is important to get the shape right as well. Fig.~\ref{fs-rreemod} shows the whole model (black line) and the two reflection components of Analysis~2; it can be compared with Fig.~5 in Ref.~\cite{fabian11}.

When we impose the Kerr metric ($\alpha_{13} = 0$), the best-fit values of some parameters are consistent with those found in Ref.~\cite{fabian11}, while others are not. However, there are several differences in the two analyses: we use different reflection models ({\sc relxill} vs {\sc reflionx}), some different input parameters (galactic dust absorption, redshift; we employ more recent measurements than~\cite{fabian11}), and there are some differences in the data reduction and analysis. Our results are consistent with those found in Ref.~\cite{kara15}, where, indeed, the differences listed above are not present.

\begin{figure*}[h]
\begin{center}
\includegraphics[type=pdf,ext=.pdf,read=.pdf,width=9.5cm]{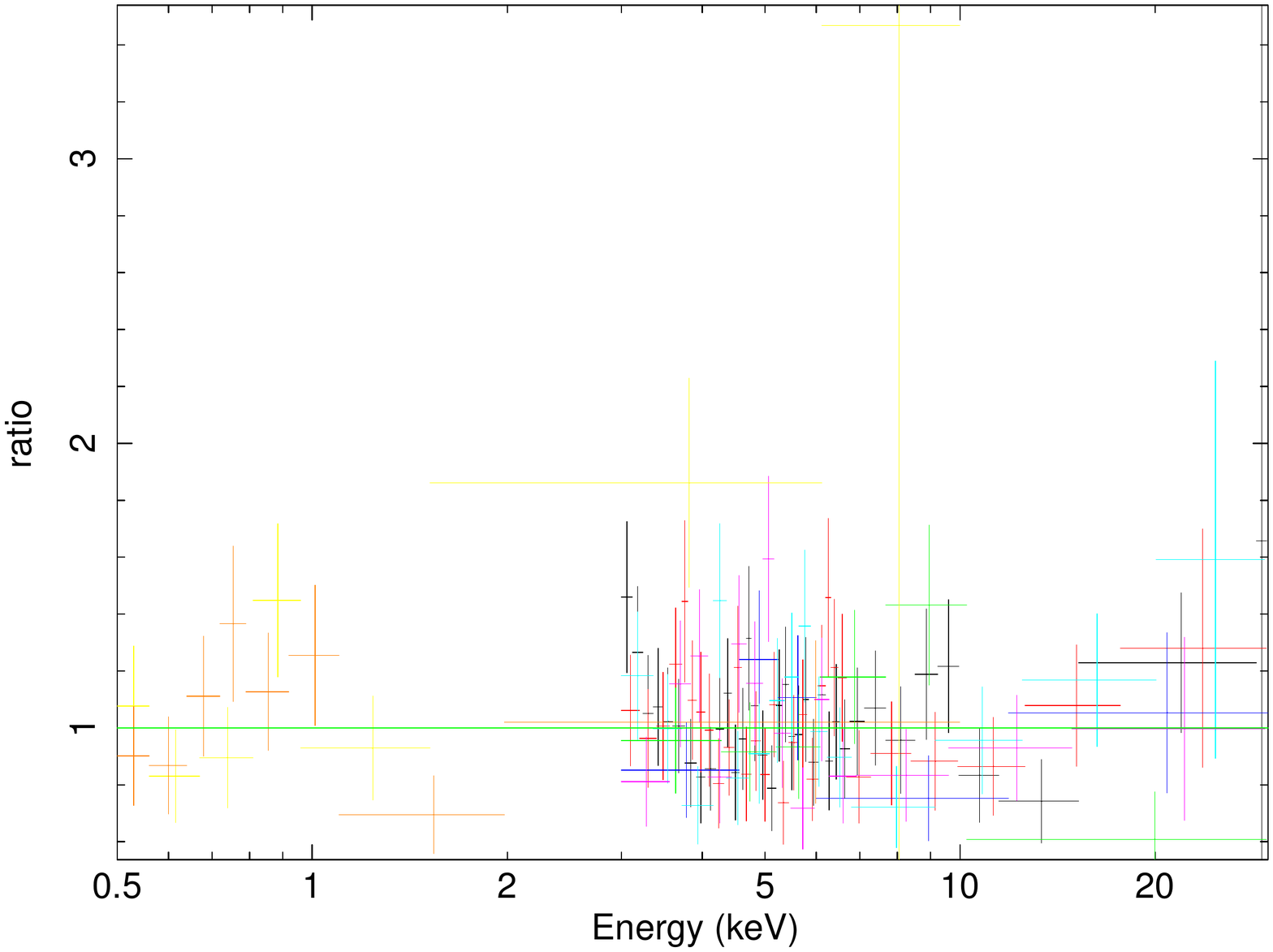}
\hspace{-1.5cm}
\includegraphics[type=pdf,ext=.pdf,read=.pdf,width=9.5cm]{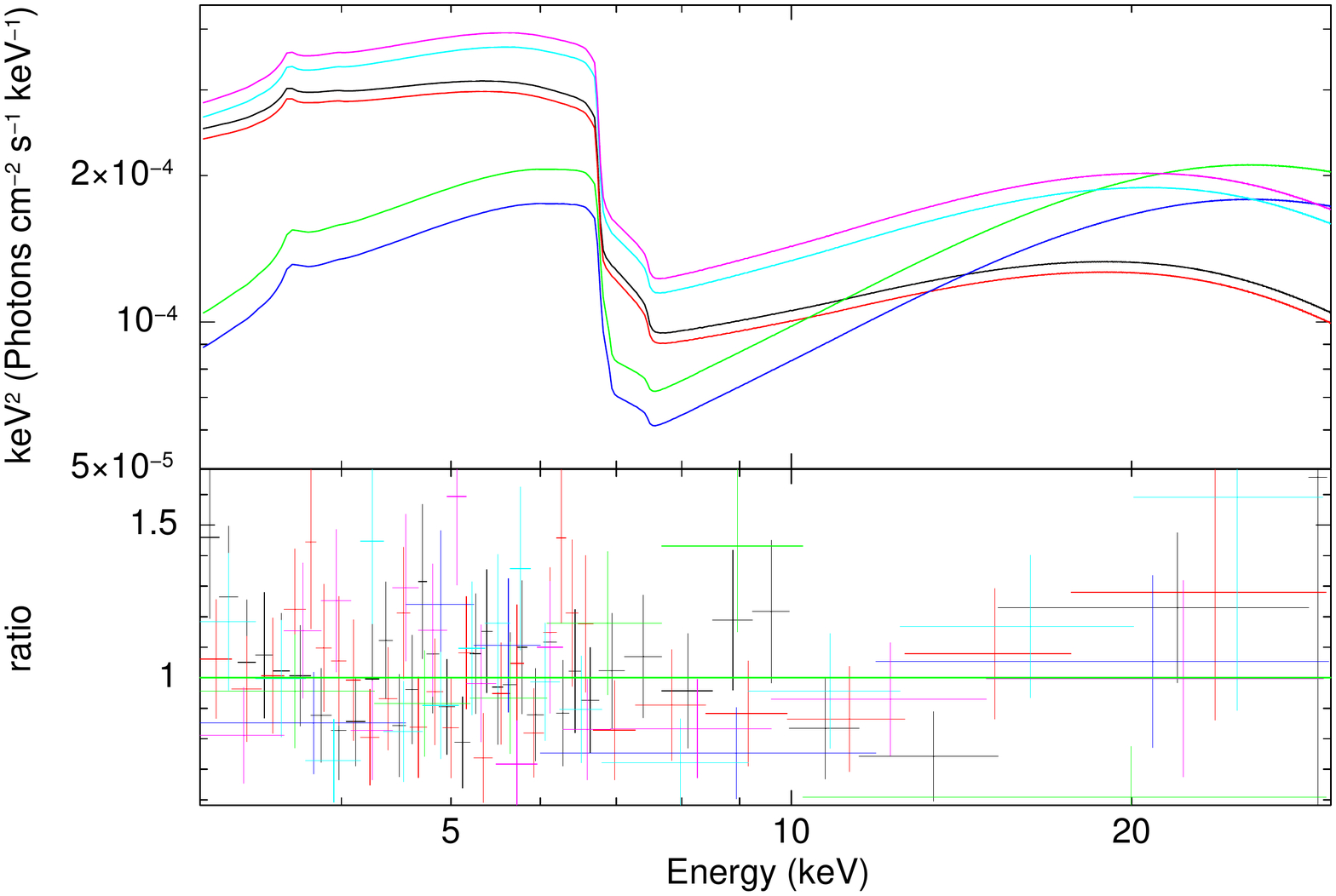}
\end{center}
\vspace{-0.2cm}
\caption{Left panel: Data-to-model ratio of the \textsl{NuSTAR}+\textsl{Swift} observations of 2014 corresponding to Analysis~3. Right panel: ${\rm energy^2 \cdot model}$ and the data to model ratio of the \textsl{NuSTAR} observations only; for every observation, we have both the FPMA and FPMB data sets. The data have been rebinned for plotting purposes only. \label{fs-nustar}}
\end{figure*}

\subsection*{\textsl{NuSTAR}+\textsl{Swift} data}

There are three separated \textsl{NuSTAR} observations of 1H0707$-$495 in 2014, with simultaneous snapshots of \textsl{Swift}/XRT. We did not include the second \textsl{Swift} observation in our analysis because it was during an anomaly period for the instrument. The \textsl{NuSTAR} data from both the FPMA and FPMB instruments were processed using nupipeline v0.4.5 with the standard filtering criteria and the NuSTAR CALDB version 20170120. For the spectra and light-curves extraction, we used the task {\it nuproduct} and we chose a circular source region of radius 40~arcsec and background region of radius 85~arcsec on the same chip. No pile-up effects were found in these \textsl{NuSTAR} observations. All spectra were binned to a minimum of 1 count before analysis. The \textsl{Swift}/XRT spectra were also extracted following the standard criteria with source region of radius 20~arcsec, using the {\it xselect} tool. The data were binned to a minimum of 1 count in order to do a simultaneous fitting with the \textsl{NuSTAR} observations. Since the signal is low, we used the Cash-statistics in the analysis of the \textsl{NuSTAR}+\textsl{Swift} data.

Fig.~\ref{fs-nustar} shows the data to model ratio of Analysis~3 \textsl{NuSTAR}+\textsl{Swift} (left panel) and the ${\rm energy^2 \cdot model}$ and the data to model ratio of the \textsl{NuSTAR} data only (right panel). The right panel in Fig.~\ref{fs-nustar} can be compared with Fig.~3 in Ref.~\cite{kara15}.

\subsection*{Systematic effects}

In our study there are several model assumptions. Some of them can indeed be important and should be investigated in a future, more systematic test of the Kerr metric with 1H0707$-$495. Others are probably not so important.

As we have already pointed out in the Letter, our results hold if the spectrum of 1H0707$-$495 is indeed reflection dominated. For example, there is now strong evidence that a powerful wind sculpts the observed spectrum around the iron line energies in IRAS13224~\cite{parker}. A similar wind has also been proposed for 1H0707$-$495~\cite{hagino}, even if for the moment most authors assume that the spectrum of this source is reflection dominated. If the wind model were correct, this would presumably have a strong impact on the best-fit. For example, in the wind model the iron abundance can be of the order of the Solar one rather than higher by a factor $\sim 10$. The aim of our Letter is to test the new reflection model and all results are based on the primary assumption that the model is in fact the correct one. For this reason we do not investigate the wind scenario, but it would be interesting and necessary to check its effects on our constraints on $a_*$ and $\alpha_{13}$.

As in any other relativistic reflection measurement, we assume that the reflecting medium has constant density (instead of the almost two phase structure produced in hydrostatic equilibrium models). At the moment it is not clear how much this assumption influences the results. However, due to the steep emissivity, most of the reflected radiation will eventually come from a very small region in the disk and actually we have only to assume that for these few gravitational radii the density is constant. Moreover, we expect that the density and ionisation partly compensate for each other, but still the assumption of a constant density reflecting medium introduces some uncertainties entering the model.

The intensity profile is quite an important ingredient and depends on the exact geometry of the corona, which is currently unknown. A simple power law is likely a crude approximation. However, it is presumably sufficient for the quality of the available data. For example, we could have considered a broken power law, which would have still been an approximation, but one a little bit more sophisticated. However, in the absence of high quality data, we cannot increase the number of parameters because we would find that we cannot get any measurement/constraint for some of them. We can expect that for the results reported in our work, the assumption of a power law is not crucial. However, precise tests of the Kerr metric with future X-ray missions definitively require a better modeling of the intensity profile.

Lastly, a crucial ingredient is usually the assumption that the inner edge of the disk is at the ISCO radius. However, for the constraints reported in the Letter we claim it is not. Indeed, if the inner edge of the disk is not at the ISCO, it should be at a larger radius (it can be at a smaller radius only for thick disks, but in this case we should not apply our thin disk model to our source). $R_{\rm in} > R_{\rm ISCO}$ leads to underestimate the spin. In our case, and in particular for Analysis~2 and Analysis~3, we only get lower bounds on the spin. This means that the inner edge of the accretion disk is already very close to the compact object. Permitting $R_{\rm in} > R_{\rm ISCO}$, we may get even stronger constraints, but in our case we cannot because the quality of the data is not good enough to constrain the spin and $R_{\rm in}$ at the same time.

\subsection*{Simulations with \textsl{Athena}}

\begin{figure*}[h]
\begin{center}
\includegraphics[type=pdf,ext=.pdf,read=.pdf,width=8.7cm]{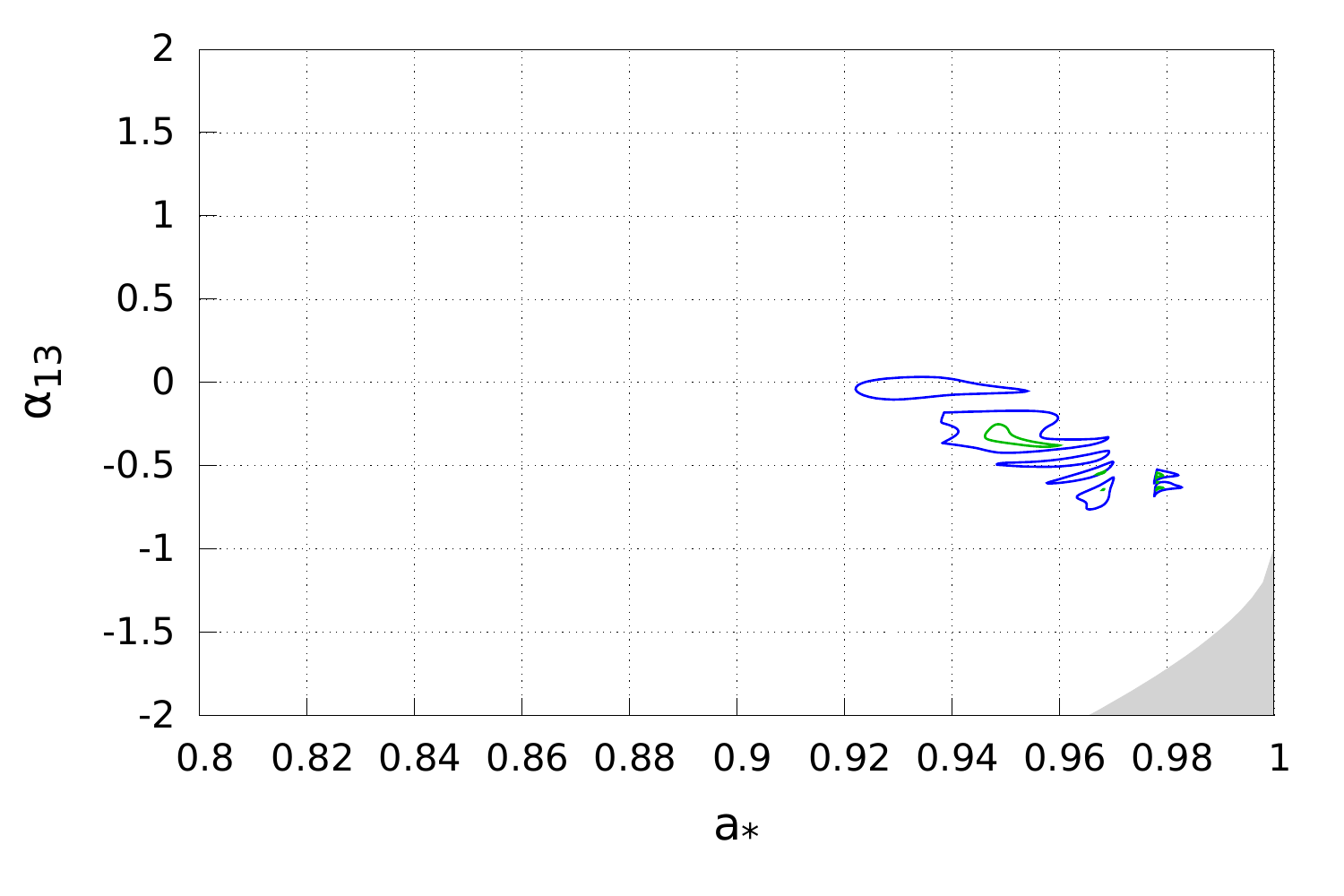}
\includegraphics[type=pdf,ext=.pdf,read=.pdf,width=8.7cm]{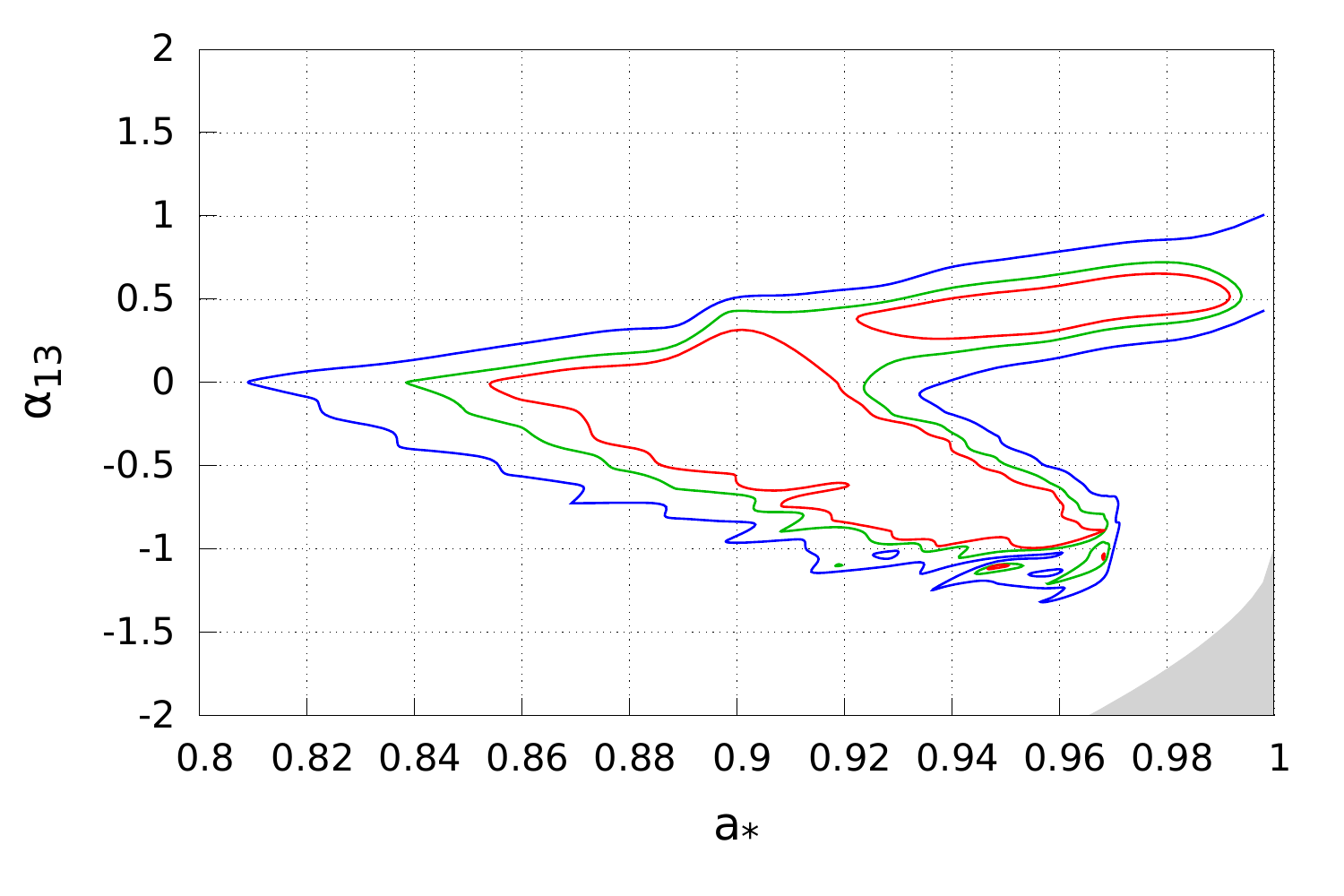}
\end{center}
\vspace{-0.5cm}
\caption{Constraints on the spin parameter $a_*$ and the Johannsen deformation parameter $\alpha_{13}$ from simulations with \textsl{X-IFU/Athena} assuming Model~1 and a Johannsen deformation parameter $\alpha_{13} = -0.5$ (left panel) and $\alpha_{13} = 0.5$ (right panel). The exposure time is $\tau = 100$~ks. The red, green, and blue lines indicate, respectively, the 68\%, 90\%, and 99\% confidence level curves for two relevant parameters. Note the difference in the range of $\alpha_{13}$ with respect to the figures shown in the Letter. \label{fs-100}}
\vspace{0.5cm}
\begin{center}
\includegraphics[type=pdf,ext=.pdf,read=.pdf,width=8.7cm]{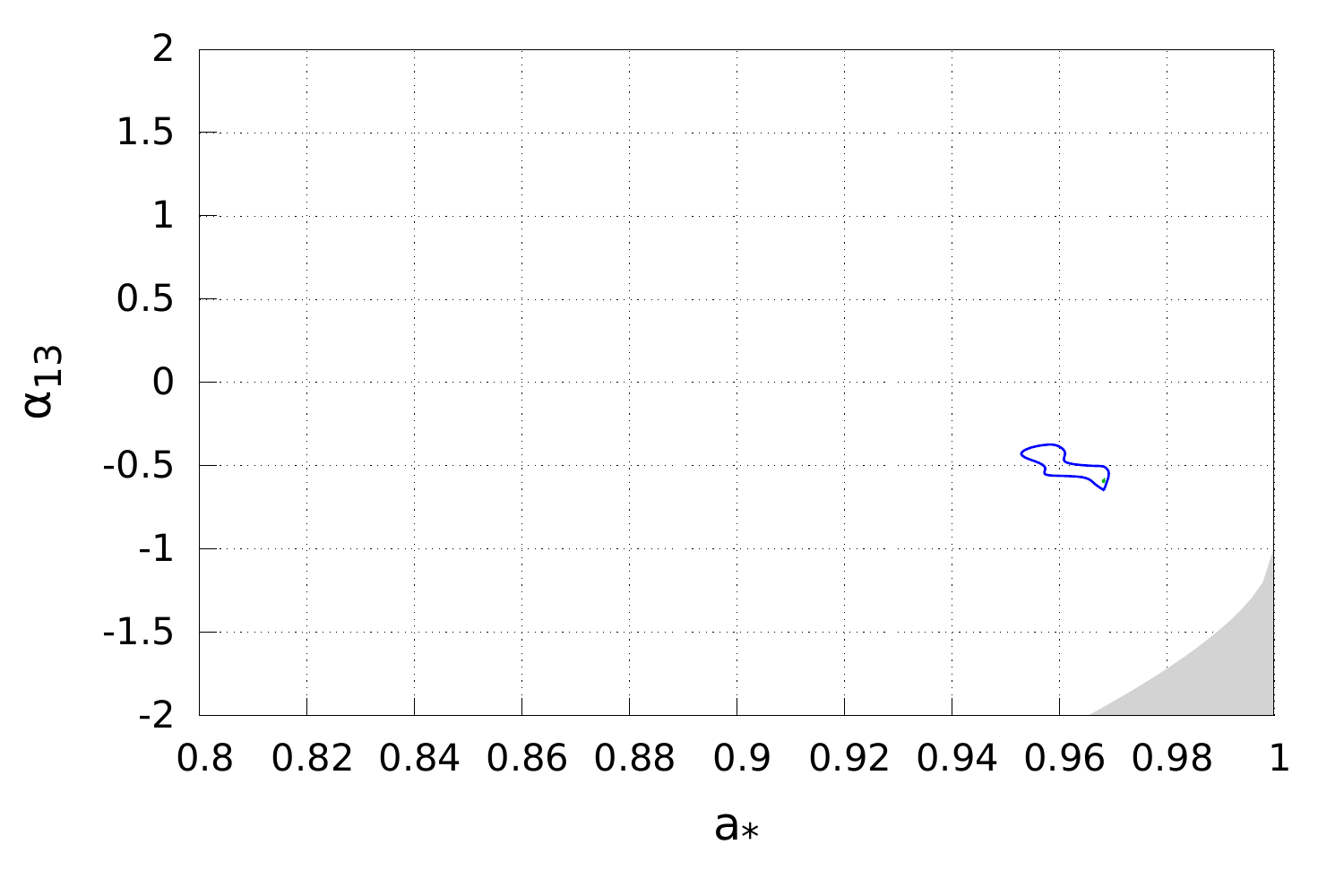}
\includegraphics[type=pdf,ext=.pdf,read=.pdf,width=8.7cm]{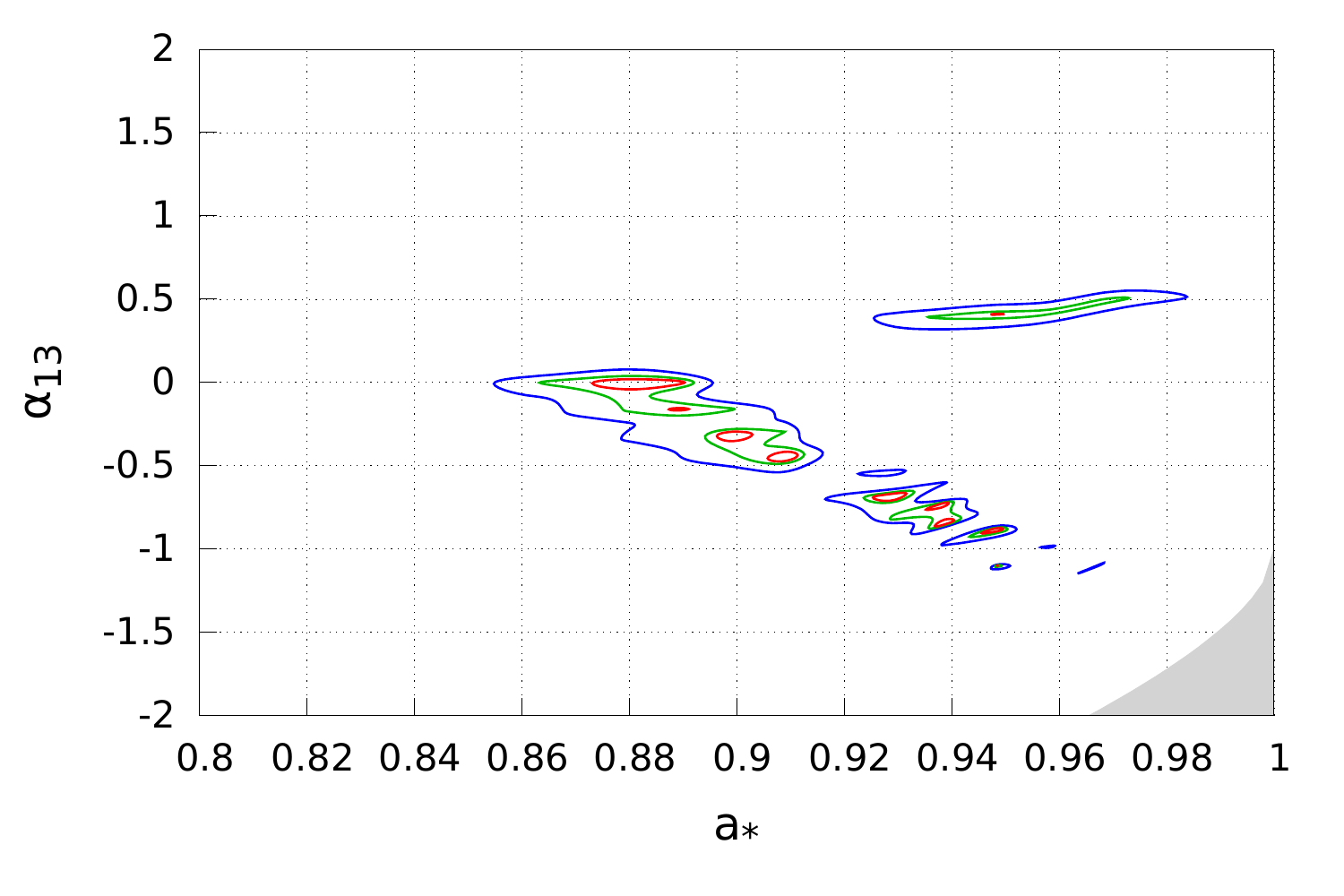}
\end{center}
\vspace{-0.5cm}
\caption{As in Fig.~\ref{fs-100} assuming an exposure time $\tau = 300$~ks. \label{fs-300}}
\end{figure*}

In order to figure out the constraining power of future X-ray missions, we perform some simulations with \textsl{X-IFU/Athena}~\cite{athena}. We assume Model~1 and plug in the best-fit values of the 2011 \textsl{XMM-Newton} data as input values for all parameters, with the exception of the deformation parameter $\alpha_{13}$. The latter is set to $\pm 0.5$ in order to see if \textsl{Athena} can potentially distinguish these metrics from the Kerr solution. Figs.~\ref{fs-100} and \ref{fs-300} show the constraints on the spin parameter $a_*$ and the Johannsen deformation parameter $\alpha_{13}$. The red, green, and blue lines indicate, respectively, the 68\%, 90\%, and 99\% confidence level curves. The assumed exposure time is 100~ks in Fig.~\ref{fs-100} and 300~ks in Fig.~\ref{fs-300}. The left panels are for $\alpha_{13} = -0.5$, the right panels are for $\alpha_{13} = 0.5$. If we perform simulations assuming Model~2 or Model~3, the constraints on $\alpha_{13}$ are stronger. The results here can be compared with those obtained in~\cite{relxill_nk}. In particular, a positive $\alpha_{13}$ solution appears to be more difficult to distinguish from a Kerr solution than a negative $\alpha_{13}$ solution. Higher exposure significantly improves detection possibility of negative $\alpha_{13}$, whereas positive $\alpha_{13}$ continues to remain degenerate with a Kerr solution. A more detailed analysis of this degeneracy and other systematic effects will appear in a forthcoming study.

\begin{thebibliography}{99}

\bibitem{hair} 
  B.~Carter,
  %``Axisymmetric Black Hole Has Only Two Degrees of Freedom,''
  Phys.\ Rev.\ Lett.\  {\bf 26}, 331 (1971);
  D.~C.~Robinson,
  %``Uniqueness of the Kerr black hole,''
  Phys.\ Rev.\ Lett.\  {\bf 34}, 905 (1975);
  P.~T.~Chrusciel, J.~L.~Costa and M.~Heusler,
  %``Stationary Black Holes: Uniqueness and Beyond,''
  Living Rev.\ Rel.\  {\bf 15}, 7 (2012).
%  [arXiv:1205.6112 [gr-qc]].  
  
\bibitem{book} 
  C.~Bambi,
  {\it Black Holes: A Laboratory for Testing Strong Gravity}
  (Springer Singapore, 2017),  doi:10.1007/978-981-10-4524-0

\bibitem{will} 
  C.~M.~Will,
  %``The Confrontation between General Relativity and Experiment,''
  Living Rev.\ Rel.\  {\bf 17}, 4 (2014)
  [arXiv:1403.7377 [gr-qc]].    
  
\bibitem{review} 
  C.~Bambi,
  %``Testing black hole candidates with electromagnetic radiation,''
  Rev.\ Mod.\ Phys.\  {\bf 89}, 025001 (2017)
  [arXiv:1509.03884 [gr-qc]].

\bibitem{ironkerr}
  A.~C.~Fabian, K.~Iwasawa, C.~S.~Reynolds and A.~J.~Young,
  %``Broad iron lines in active galactic nuclei,''
  Publ.\ Astron.\ Soc.\ Pac.\  {\bf 112}, 1145 (2000)
  [astro-ph/0004366];
  L.~W.~Brenneman and C.~S.~Reynolds,
  %``Constraining Black Hole Spin Via X-ray Spectroscopy,''
  Astrophys.\ J.\  {\bf 652}, 1028 (2006)
  [astro-ph/0608502];
  C.~S.~Reynolds,
  %``Measuring Black Hole Spin using X-ray Reflection Spectroscopy,''
  Space Sci.\ Rev.\  {\bf 183}, 277 (2014)
  [arXiv:1302.3260 [astro-ph.HE]]. 

\bibitem{iron}
  J.~Schee and Z.~Stuchlik,
  %``Profiles of emission lines generated by rings orbiting braneworld Kerr black holes,''
  Gen.\ Rel.\ Grav.\  {\bf 41}, 1795 (2009)
  [arXiv:0812.3017 [astro-ph]];
  T.~Johannsen and D.~Psaltis,
  %``Testing the No-Hair Theorem with Observations in the Electromagnetic Spectrum. IV. Relativistically Broadened Iron Lines,''
  Astrophys.\ J.\  {\bf 773}, 57 (2013)
  [arXiv:1202.6069 [astro-ph.HE]];
  C.~Bambi,
  %``Testing the space-time geometry around black hole candidates with the analysis of the broad K$\alpha$ iron line,''
  Phys.\ Rev.\ D {\bf 87}, 023007 (2013)
  [arXiv:1211.2513 [gr-qc]];
  J.~Jiang, C.~Bambi and J.~F.~Steiner,
  %``Testing the Kerr Nature of Black Hole Candidates using Iron Line Spectra in the CPR Framework,''
  Astrophys.\ J.\  {\bf 811}, 130 (2015)
  [arXiv:1504.01970 [gr-qc]];
  C.~Bambi, J.~Jiang and J.~F.~Steiner,
  %``Testing the no-hair theorem with the continuum-fitting and the iron line methods: a short review,''
  Class.\ Quant.\ Grav.\  {\bf 33}, 064001 (2016)
  [arXiv:1511.07587 [gr-qc]];
  J.~K.~Hoormann, B.~Beheshtipour and H.~Krawczynski,
  %``Testing general relativityÕs no-hair theorem with x-ray observations of black holes,''
  Phys.\ Rev.\ D {\bf 93}, 044020 (2016)
  [arXiv:1601.02055 [astro-ph.HE]];  
  M.~Zhou, A.~Cardenas-Avendano, C.~Bambi, B.~Kleihaus and J.~Kunz,
  %``Search for astrophysical rotating Ellis wormholes with X-ray reflection spectroscopy,''
  Phys.\ Rev.\ D {\bf 94}, 024036 (2016)
  [arXiv:1603.07448 [gr-qc]];  
  Y.~Ni, M.~Zhou, A.~Cardenas-Avendano, C.~Bambi, C.~A.~R.~Herdeiro and E.~Radu,
  %``Iron K$\alpha$ line of Kerr black holes with scalar hair,''
  JCAP {\bf 1607}, 049 (2016)
  [arXiv:1606.04654 [gr-qc]];
  C.~Bambi, Z.~Cao and L.~Modesto,
  %``Testing conformal gravity with astrophysical black holes,''
  Phys.\ Rev.\ D {\bf 95}, 064006 (2017)
  [arXiv:1701.00226 [gr-qc]].  

\bibitem{relxill} 
  T.~Dauser, J.~Garcia, J.~Wilms, M.~Bock, L.~W.~Brenneman, M.~Falanga, K.~Fukumura and C.~S.~Reynolds,
  %``Irradiation of an Accretion Disc by a Jet: General Properties and Implications for Spin Measurements of Black Holes,''
  Mon.\ Not.\ Roy.\ Astron.\ Soc.\  {\bf 430}, 1694 (2013)
  [arXiv:1301.4922 [astro-ph.HE]];
  J.~Garcia, T.~Dauser, C.~S.~Reynolds, T.~R.~Kallman, J.~E.~McClintock, J.~Wilms and W.~Eikmann,
  %``X-ray reflected spectra from accretion disk models. III. A complete grid of ionized reflection calculations,''
  Astrophys.\ J.\  {\bf 768}, 146 (2013)
  [arXiv:1303.2112 [astro-ph.HE]]; 
  J.~Garc'a {\it et al.},
  %``Improved Reflection Models of Black-Hole Accretion Disks: Treating the Angular Distribution of X-rays,''
  Astrophys.\ J.\  {\bf 782}, 76 (2014)
  [arXiv:1312.3231 [astro-ph.HE]].  
  
\bibitem{relxill_nk} 
  C.~Bambi, A.~Cardenas-Avendano, T.~Dauser, J.~A.~Garcia and S.~Nampalliwar,
  %``Testing the Kerr black hole hypothesis using X-ray reflection spectroscopy,''
  Astrophys.\ J.\  {\bf 842}, 76 (2017)
  [arXiv:1607.00596 [gr-qc]].  
  
\bibitem{corona}
  G.~Matt, G.~C.~Perola and L.~Piro,
  Astron.\ Astrophys.\  {\bf 247}, 25 (1991);
  A.~Martocchia and G.~Matt,
  Mon.\ Not.\ Roy.\ Astron.\ Soc.\  {\bf 282}, L53 (1996).  

\bibitem{j-m} 
  T.~Johannsen,
  %``Regular Black Hole Metric with Three Constants of Motion,''
  Phys.\ Rev.\ D {\bf 88}, 044002 (2013)
  [arXiv:1501.02809 [gr-qc]]. 
  
\bibitem{0707} 
  A.~C.~Fabian {\it et al.},
  %``Broad line emission from iron K- and L-shell transitions in the active galaxy 1H?0707-495,''
  Nature {\bf 459}, 540 (2009)
  [arXiv:0905.4383 [astro-ph.HE]];
  A.~Zoghbi, A.~Fabian, P.~Uttley, G.~Miniutti, L.~Gallo, C.~Reynolds, J.~Miller and G.~Ponti,
  %``Broad iron L-line and X-ray reverberation in 1H0707-495,''
  Mon.\ Not.\ Roy.\ Astron.\ Soc.\  {\bf 401}, 2419 (2010)
  [arXiv:0910.0367 [astro-ph.HE]].  
  
\bibitem{fabian11} 
  A.~C.~Fabian {\it et al.},
  %``1H0707-495 in 2011: An X-ray source within a gravitational radius of the event horizon,''
ÊÊMon.\ Not.\ Roy.\ Astron.\ Soc.\  {\bf 419}, 116 (2012)
ÊÊ[arXiv:1108.5988 [astro-ph.HE]].   
  
\bibitem{0707td12} 
  T.~Dauser {\it et al.},
  %``Spectral Analysis of 1H0707-495 with XMM-Newton,''
  Mon.\ Not.\ Roy.\ Astron.\ Soc.\  {\bf 422}, 1914 (2012)
  [arXiv:1112.1796 [astro-ph.HE]].  
  
\bibitem{kara15} 
  E.~Kara {\it et al.},
  %``The Compton hump and variable blue wing in the extreme low-flux NuSTAR observations of 1H0707?495,''
  Mon.\ Not.\ Roy.\ Astron.\ Soc.\  {\bf 449}, 234 (2015)
  [arXiv:1501.06849 [astro-ph.HE]].
  
\bibitem{parker} 
  M.~L.~Parker {\it et al.},
  %``The response of relativistic outflowing gas to the inner accretion disk of a black hole,''
  Nature {\bf 543}, 83 (2017)
  [arXiv:1703.00071 [astro-ph.HE]].

\bibitem{hagino} 
  K.~Hagino, H.~Odaka, C.~Done, R.~Tomaru, S.~Watanabe and T.~Takahashi,
  %``A disc wind interpretation of the strong Fe K$\alpha$ features in 1H 0707?495,''
  Mon.\ Not.\ Roy.\ Astron.\ Soc.\  {\bf 461}, 3954 (2016)
  [arXiv:1509.05645 [astro-ph.HE]];
  M.~L.~Parker {\it et al.} (in preparation).  
  
\bibitem{lohfink} 
  A.~M.~Lohfink, C.~S.~Reynolds, J.~M.~Miller, L.~W.~Brenneman, R.~F.~Mushotzky, M.~A.~Nowak and A.~C.~Fabian,
  %``The Black Hole Spin and Soft X-ray Excess of the Luminous Seyfert Galaxy Fairall 9,''
  Astrophys.\ J.\  {\bf 758}, 67 (2012)
  [arXiv:1209.0468 [astro-ph.HE]].  
  
\bibitem{wilkins14}  
  D.~R.~Wilkins, E.~Kara, A.~C.~Fabian and L.~C.~Gallo,
  %``Caught in the act: measuring the changes in the corona that cause the extreme variability of 1H 0707?495,''
  Mon.\ Not.\ Roy.\ Astron.\ Soc.\  {\bf 443}, 2746 (2014)
 [arXiv:1406.6658 [astro-ph.HE]].
  
\bibitem{snzhang} 
  S.~N.~Zhang {\it et al.} [eXTP Collaboration],
  %``eXTP -- enhanced X-ray Timing and Polarimetry Mission,''
  Proc.\ SPIE Int.\ Soc.\ Opt.\ Eng.\  {\bf 9905}, 99051Q (2016)
  [arXiv:1607.08823 [astro-ph.IM]]. 

\bibitem{athena}
  http://www.the-athena-x-ray-observatory.eu/

\end{thebibliography}

\begin{thebibliography}{99}
   

\bibitem{j-m} 
  T.~Johannsen,
  %``Regular Black Hole Metric with Three Constants of Motion,''
  Phys.\ Rev.\ D {\bf 88}, 044002 (2013)
  [arXiv:1501.02809 [gr-qc]]. 
 
\bibitem{relxill_nk} 
  C.~Bambi, A.~Cardenas-Avendano, T.~Dauser, J.~A.~Garcia and S.~Nampalliwar,
  %``Testing the Kerr black hole hypothesis using X-ray reflection spectroscopy,''
  Astrophys.\ J.\  {\bf 842}, 76 (2017)
  [arXiv:1607.00596 [gr-qc]]. 
  
\bibitem{fabian11} 
  A.~C.~Fabian {\it et al.},
  %``1H0707-495 in 2011: An X-ray source within a gravitational radius of the event horizon,''
ÊÊMon.\ Not.\ Roy.\ Astron.\ Soc.\  {\bf 419}, 116 (2012)
ÊÊ[arXiv:1108.5988 [astro-ph.HE]].    
  
\bibitem{kara15} 
  E.~Kara {\it et al.},
  %``The Compton hump and variable blue wing in the extreme low-flux NuSTAR observations of 1H0707?495,''
  Mon.\ Not.\ Roy.\ Astron.\ Soc.\  {\bf 449}, 234 (2015)
  [arXiv:1501.06849 [astro-ph.HE]].
  
\bibitem{parker} 
  M.~L.~Parker {\it et al.},
  %``The response of relativistic outflowing gas to the inner accretion disk of a black hole,''
  Nature {\bf 543}, 83 (2017)
  [arXiv:1703.00071 [astro-ph.HE]].

\bibitem{hagino} 
  K.~Hagino, H.~Odaka, C.~Done, R.~Tomaru, S.~Watanabe and T.~Takahashi,
  %``A disc wind interpretation of the strong Fe K$\alpha$ features in 1H 0707?495,''
  Mon.\ Not.\ Roy.\ Astron.\ Soc.\  {\bf 461}, 3954 (2016)
  [arXiv:1509.05645 [astro-ph.HE]];
  M.~L.~Parker {\it et al.} (in preparation).  

\bibitem{athena}
  http://www.the-athena-x-ray-observatory.eu/
  
\end{thebibliography}
\end{document}